\documentclass[12pt,preprint]{aastex}
\usepackage{psfig}

\newcommand{\andix}{Andromeda {\sc IX}}

\newcommand{\msun}{${\rm M_{\sun}}$}

\def\kms{{\rm\,km\,s^{-1}}}

\def\kpc{{\rm\,kpc}}

\def\msun{{\rm\,M_\odot}}
\def\lsun{{\rm\,L_\odot}}

\def\pc{{\rm\,pc}}

\slugcomment{{\sc to appear in ApJ Letters: }  }

\begin{document}

\title{A Keck/DEIMOS Kinematic Study of Andromeda IX: dark matter on the 
smallest galactic scales}

\author{Scott C.\ Chapman\altaffilmark{1}, 
Rodrigo\ Ibata\altaffilmark{2}, Geraint\ F.\ Lewis\altaffilmark{3},   
Annette M.\ N.\ Ferguson\altaffilmark{4}, 
Mike\ Irwin\altaffilmark{5}, Alan McConnachie\altaffilmark{5}, 
Nial Tanvir\altaffilmark{6}
}
\altaffiltext{1}{California Institute of Technology, Pasadena, CA\,91125;
 \texttt{schapman@astro.caltech.edu}}
\altaffiltext{2}{
Observatoire de Strasbourg, 11, rue de l'Universit\'e, F-67000, Strasbourg,
France} 
\altaffiltext{3}{
Institute of Astronomy, School of Physics, A29, University of Sydney, NSW
2006, Australia}
\altaffiltext{4}{Institute for Astronomy, University of Edinburgh, Royal Observatory, Blackford Hill, Edinburgh, EH9 3HJ, U.K.}	  		
\altaffiltext{5}{
Institute of Astronomy, Madingley Road, Cambridge, CB3 0HA, U.K.}
\altaffiltext{6}{
Centre for Astrophysics Research, Univ. of Hertfordshire, Hatfield, AL10 9AB, UK}

\begin{abstract}
We present the  results of a kinematic survey  of the dwarf spheroidal
satellite  of M31,  \andix, which  appears  to be  the lowest  surface
brightness and also  the faintest galaxy (${\rm M_{V}}  = -8.3$) found
to date.   Using Keck/DEIMOS spectroscopic data, we  have measured its
velocity   relative  to   M31,  its   velocity  dispersion,   and  its
metallicity.    It   exhibits   a  significant   velocity   dispersion
$\sigma_{v}   =  6.8^{+3.0}_{-2.0}\kms$,  which   coupled  with   the  low
luminosity implies a very high mass to $V$-band light ratio, $M/L \sim
93^{+120}_{-50}\msun/\lsun$   ($M/L    >   17\msun/\lsun$   at   99\%
confidence). Unless  strong tidal  forces have perturbed  this system,
this smallest of galaxies is a highly dark matter dominated system.
\end{abstract}


\section{Introduction}
\label{txt:intro}

Hierarchical structure formation models, such as $\Lambda$CDM, predict
that large spiral galaxies like  the Milky Way (MW) or Andromeda (M31)
arise from  successive mergers of  small galaxies and from  the smooth
accretion of gas.   At late times, when a dominant  central mass is in
place, the main mode of mass acquisition is via the cannibalization of
low mass  satellites which fall into the  gravitational potential well of
the massive host.

These  $\Lambda$CDM   models  are  highly   successful  at  explaining
observations  on  large  scales.    However,  the  models  predict  an
overabundance of low-mass dark subhalos, which is inconsistent by 1--2
orders  of  magnitude  with  the  number of  observed  dwarf  galaxies
(\citealt{klypin99, moore99,  benson02a}).  Large uncertainties remain
on the level of disagreement  between model and observation because of
the difficulty in constraining the  faint end of the galaxy luminosity
function;  the  low surface  brightnesses  expected  for the  faintest
galaxies   ($\mu_{V}  \gtrsim   26\,  {\rm   mag~arcsec}^{-2}$;  e.g.,
\citealt{caldwell99, benson02b})  imply that ground-based  surveys are
likely incomplete  even in nearby  galaxy groups.  From  a theoretical
perspective, star  formation in low  mass subsystems would have  to be
suppressed  (e.g., photoionization  in  the early  universe) to  bring
models  and  observations into  closer  accord.   The  result of  this
hydro-dynamical modification would be  a shallower faint-end slope for
the $z\sim0$ galaxy  luminosity function (e.g., \citealt{somerville02,
benson02b,willman04}).   
A critical prediction  of this  solution is  that dwarf
galaxies should  then be  embedded in much  larger, more  massive dark
subhalos \citep{stoehr02}.

Observational  progress can be  made in  the case  of the  Local Group
(LG), where galaxies can be  resolved into stars, thereby probing much
fainter   effective    surface   brightness   limits   \citep{ibata01,
ferguson02, lewis04,  willman05}.  Several recent  and ongoing surveys
are yielding  exciting detections  of low surface  brightness features
and new  dwarf galaxy candidates  around our nearest  giant neighbour,
M31   (e.g.   \citealt{ibata01,   ferguson02,   zucker04a,  zucker04b,
mcconnachie04}).  Of  particular interest is the  system And~IX, which
was   first  reported   by   the  Sloan   Digital   Sky  Survey   team
\citep{zucker04a}, and is readily visible in previously-published maps
from the M31 INT Wide-Field Camera Survey \citep{ferguson02}.  \andix\
is unique in that it is  the lowest surface brightness galaxy found to
date  ($\mu_{V,0}  \sim 26.8\,  {\rm  mag~arcsec}^{-2}$),  and at  the
distance  estimated from  the position  of the  tip of  the  red giant
branch ($I=20.50\pm0.03$, $M_I^{TRGB}=4.065$,  $(m - M)_0 \sim 24.42$)
of  $765\pm24 \pc$  \citep{mcconnachie05}, \andix\  would also  be the
faintest  galaxy known  (${\rm M_{V}}  = -8.3$).   It is  therefore an
extremely important  target for further  study, with the  potential to
place strong  constraints on both  the LG luminosity function  and the
physics of galaxy  formation at the smallest scales.   In this letter,
we  present the results  of a  kinematic survey  of the  \andix\ dwarf
spheroidal satellite of M31. We have assumed for this paper a distance
to   M31   of   $785\pm25$\,kpc   ($(m   -   M)_0   =   24.47\pm0.07$;
\citep{mcconnachie05}).

\section{Properties of \andix}

\andix\ reveals itself in the INT WFC survey data as an enhancement of
metal  poor RGB  stars  at a  location  of $1.8^\circ$  to  the E  and
$1.9^\circ$  to  the  N of  the  nucleus  of  M31.  Figure~1  shows  a
color-magnitude  diagram  (CMD)  centered  on  \andix\  and  a  nearby
comparison  region of equivalent  area.  The  excess red  giant branch
(RGB) stars on  the blueward side of the general M31  RGB locus form a
well-defined  \andix\  locus with  an  apparent  RGB  tip at  $I=20.5$
consistent with the  distance of M31, and if  representative of an old
stellar  population, $[Fe/H]  \sim -1.5$  \citep{mcconnachie05}.  This
metallicity value could be a slight overestimate given the presence of
contaminating stars from  the halo of M31; indeed  the deeper study by
\cite{harbeck} found  that the  \andix\ population is  more metal-poor
($[Fe/H] \sim -2.0$).

\begin{figure}
\centerline{
\psfig{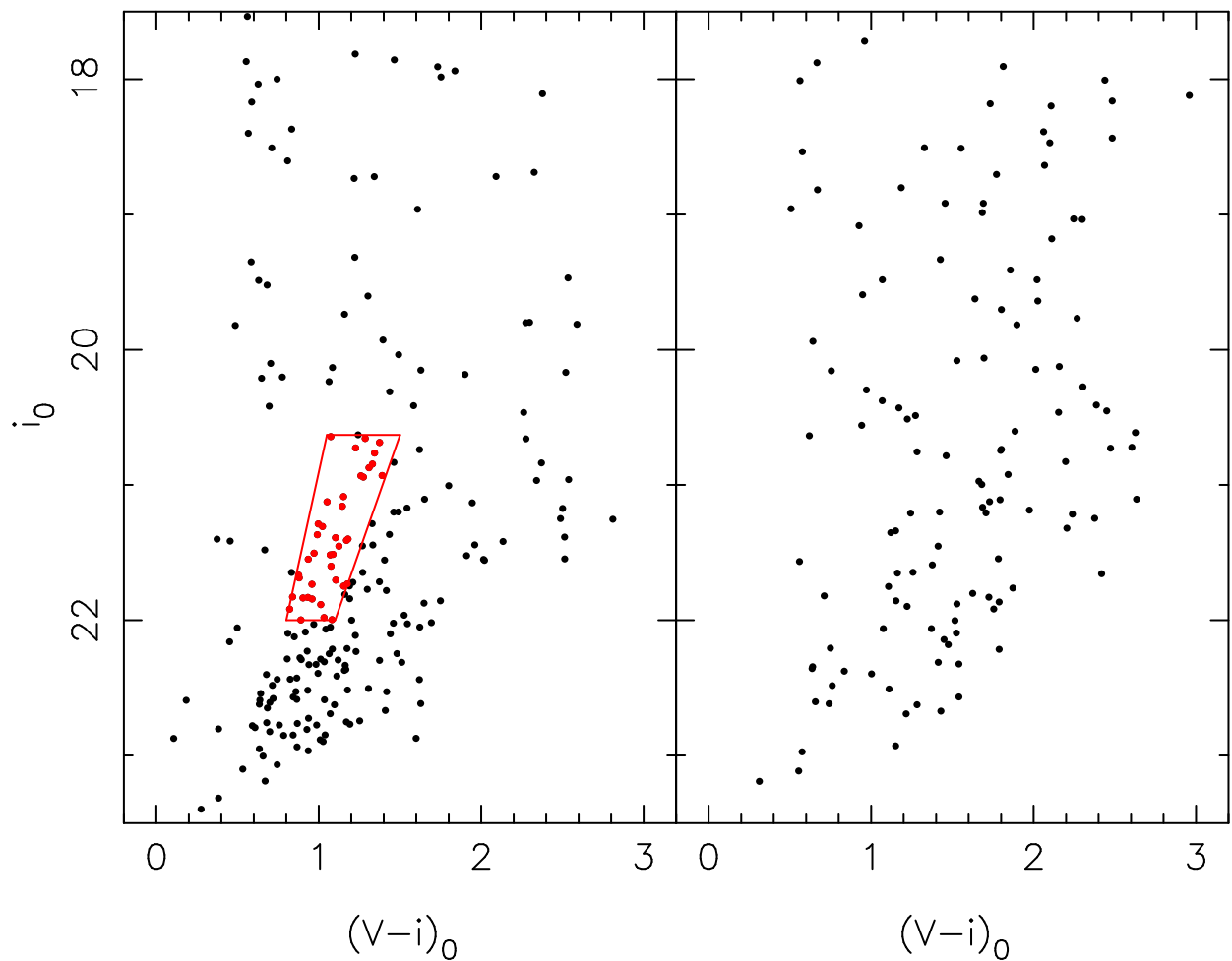}
}
\caption{Left  panel:  the   color-magnitude  diagram  from  a  region
centered on \andix\ with  radius $2.5$\arcmin. The quadrilateral shows
the adopted CMD  selection box, designed to select  \andix\ RGB stars,
which  are blue  and  metal-poor, for  spectroscopic follow-up.  Right
panel: a comparison region of the same size $10$\arcmin to the North.}
\end{figure}

Although  sparsely  populated, the  radial  profile,  computed as  the
median counts in circular annuli centered on $00^h 52^m 52^s +43^\circ
12' 00''$,  is well defined  and is shown background-corrected  in the
right-hand  panel of Figure~2.   Exponential and  Plummer-law profiles
provide adequate fits, with the  scale radius from the exponential fit
being $r_s=1.4  \pm 0.3$\arcmin\ (corresponding  to $r_s = 315  \pm 65
\pc$; half-light radius  $r_h = 530 \pm 110  \pc$), a value comparable
to  the  other Andromedan  dSph  companions \citep{mateo,  caldwell99,
mcconnachie05b}.  The  error estimates include the  uncertainty in the
background subtraction.

The luminosity  of the  dwarf galaxy is  constrained by  measuring the
integrated flux,  or surface brightness,  distribution. The processing
procedure \citep{irwin04}  is relatively straightforward:  first the
existing  derived object catalogues  are used  to define  a ``bright''
foreground star component,  one magnitude above the RGB  tip (to allow
for potential AGB  stars); a circular aperture is  excised around each
foreground  star  and the  flux  within set  to  the  local sky  level
interpolated  from  a whole-frame  background  map  (the  size of  the
aperture is  the maximum of 4  times the catalog-recorded  area at the
detection isophote  or a  diameter 4 times  the derived  FWHM seeing);
each frame  is then  rebinned on  a $3 \times  3$ grid  to effectively
create $1\arcsec$  pixels; the binned  image is then  further smoothed
using a 2D Gaussian filter of FWHM $5\arcsec$.

The  result of  this is  to  produce a  coarsely-sampled smooth  image
containing both  the resolved  and unresolved light  contribution from
the dwarf. Elliptical apertures centered  on the dwarf are then placed
over the  mosaic. The background  is determined by robustly  fitting a
smoothly-varying  surface over  the  whole mosaic  area.  The  central
surface  brightness can  then  be trivally  measured  by deriving  the
radial profile, defined as  the background-corrected median flux value
within elliptical annuli.  The variation in the flux from the multiple
background measures gives  a good indication of the  flux error, which
is,  of course, dominated  by systematic  fluctuations rather  than by
random noise.  Since the  systematics are dominated by random residual
foreground star halos and  scattered light from bright stars, possibly
just outside  the field-of-view,  we chose to  mitigate the  effect by
defining the size  of the elliptical aperture to  be equivalent to the
geometric half-light  radius derived from the  number density profile.
The estimated total flux is  then scaled to allow for this correction.
In this  way we  derive ${\rm  M_{V}} = -8.3  \pm 0.2$  (corrected for
extinction), in  agreement with the  estimate of ${\rm M_{V}}  = -8.3$
from  \cite{zucker04a};  and  measure  a  central  surface  brightness
$\Sigma_V = 26.5 \pm 0.3 \, {\rm mag \, arcsec^{-2}}$.

\begin{figure}
\centerline{
\psfig{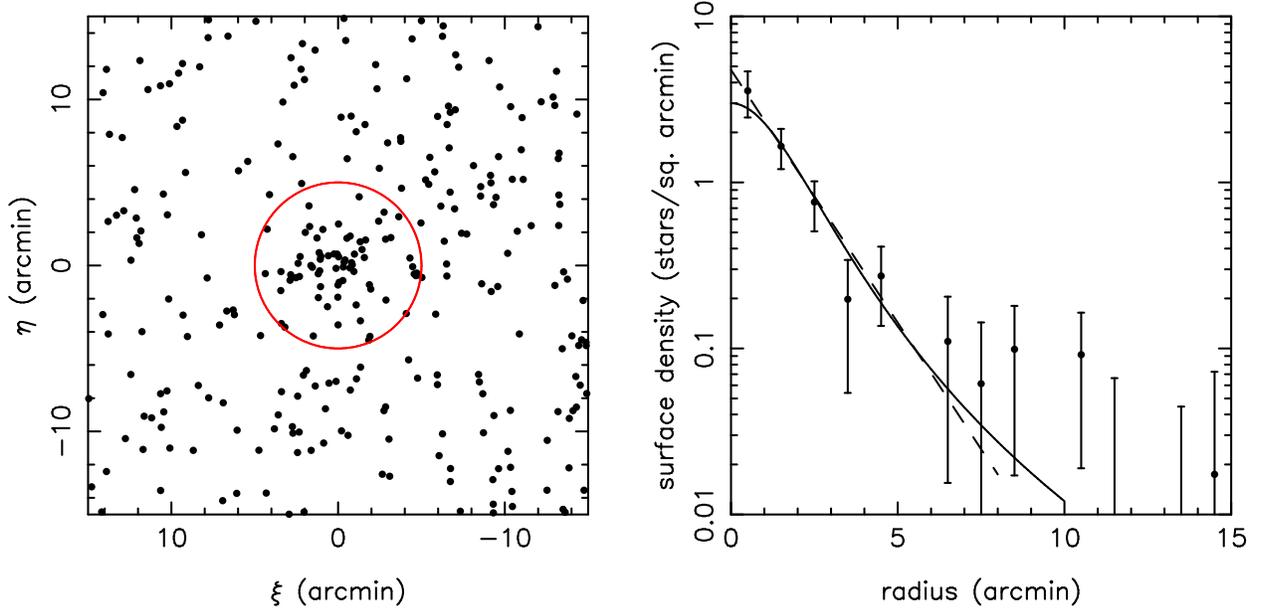}
}
\vspace{6pt}
\caption{The  left panel shows  the positions  of the  stars in  a $15
\times 15'$ region  around the center of \andix\ that  fall in the RGB
selection box of  Figure~1. The $5'$ radius circle  used for selecting
radial velocity members is  also displayed. The right-hand panel shows
the  surface density  profile of  these  RGB stars  with a  background
component  removed.   This  background  was modelled  with  a  sloping
component that is  a function of the radial distance  to the center of
M31; at the position of \andix\ the background has a number density of
$0.26 \pm  0.08$ arcmin$^{-2}$.  The  overlaid profiles are  a Plummer
model (solid  line) with central  density 3.0 arcmin$^{-2}$  and scale
size  $2.6$\arcmin\  and an  exponential  profile  (dashed line)  with
central density 4.8 arcmin$^{-2}$ and scale size $1.4$\arcmin.}
\end{figure}

\section{Spectroscopic Observations and Data Analysis}

Spectra for \andix\ were obtained  with the DEIMOS spectrograph on the
Keck2 telescope  on the  night of Sept.\  13, 2004,  under photometric
conditions  and seeing  from 0.5\arcsec--0.8\arcsec.  We  employed the
1200~${\rm l/mm}$ grating covering the wavelength range 6400--9000\AA,
with a spectral resolution of $\sim$1.5\AA. The observations pioneered
a  new approach  with  DEIMOS: using  a  `fibre-hole' 0.7$''$  slitlet
approach packing $\sim$600 holes per mask.

For the  \andix\ field,  we constructed a  `fibre-hole' mask  with 623
holes  assigned  within  the  $16'  \times  5'$  DEIMOS  field.   This
`fibre-hole'  approach  adopted  in  Sept.\  2004 proved  to  be  very
successful, giving Poisson-limited  sky-subtraction (down to $i=21.5$)
by assigning holes  to monitor the sky spectrum.   Stars were selected
for  observation primarily  within a  color-magnitude box  designed to
choose red-giant  branch stars  from the 5\arcmin\  region surrounding
\andix. This  color-magnitude selection box is  displayed in Figure~1.
We then  let the DEIMOS configuration program  randomly choose objects
as fillers with I-band magnitudes between  ${\rm 20.5 < I < 22.0}$ and
colors  ${\rm  1.0 <  V-i  <  4.0}$.  Both metal-poor  and  metal-rich
populations  will  be  present  in  this  selection.   The  fiber-hole
diameters were milled  at $0.7$\arcsec\ to match the  median seeing. A
total of 289 stars and 334 sky slits were assigned in this low-density
field in the M31 outer halo.

The \andix\ mask was observed  for three integrations of 20\,min each,
nodding to blank  sky for 5\,min between each  exposure to ensure that
an adequate sky model  could be constructed.  The spectroscopic images
were processed  and combined using the pipeline  software developed by
our group.  This software  debiasses, performs a flat-field, extracts,
wavelength-calibrates  and  sky-subtracts  the  spectra.   The  radial
velocities of the stars were  then measured with respect to a Gaussian
model  of  the  CaII  triplet  lines  (similar  to  the  technique  of
\cite{wilkinson}).  By fitting the  three Ca triplet lines separately,
an  estimate of  the radial  velocity accuracy  was obtained  for each
radial   velocity   measurement.    The  measurements   have   typical
uncertainties of  $5\kms$ to $10\kms$.  A sample of 138  stars yielded
a continuum S/N$>$10 and
cross-correlations with velocity  uncertainties of less than $15\kms$,
and  we include only  these stars  in the  subsequent analysis;  18 of
these 138 stars lie in the \andix\ color selection window.

The  measured radial  velocities in  the  \andix\ field  are shown  in
Figure~3  as a  function of  radial  distance from  the galaxy,  where
filled circles  show the color-selected stars, while  open circles are
stars  outside the CMD  selection box.   A strong  foreground Galactic
component is present at heliocentric velocities $v > -150\kms$, though
as we show  in a companion paper \citep{ibata05},  below this velocity
the Galactic contamination  is small.  Close to the  center of \andix\
there is a small but clear  kinematic grouping of stars which also lie
on the \andix\ RGB.  We first define an inner sample to be the 5 stars
observed   within   a   radius   of   one   exponential   scale-length
($1.4$\arcmin); this  sample has mean of $v=-210.1\kms$  and an r.m.s.
dispersion of  $\sigma_v=5.2\kms$, which is tiny compared  to the huge
velocity spread of stars beyond $5$\arcmin, indicating that the sample
is highly unlikely to be  contaminated. Indeed, outside of a radius of
$5\arcmin$ not a single RGB-selected  star is found within $80\kms$ of
this mean.

An  inspection  of  Figure~3  shows  that all  stars  within  the  RGB
selection box in  a radius of $5\arcmin$ have  radial velocities close
to  the mean  of  the inner  sample.   A fit  of a  maximum-likelihood
Gaussian model given these data (and their uncertainties), is shown on
the left-hand panel of Figure~4  (thin black lines), where we display
the likelihood  contours as a  function of mean velocity  and velocity
dispersion.   The most  likely  values  are $v  =  -219\pm 4\kms$  and
$\sigma_v   =  12.9^{+4}_{-2}\kms$.    However,  a   single   star  at
$1.9\arcmin$ stands out  in this sample of 11 objects  as it is offset
by $-44.6\kms$ from the mean velocity of the inner sample.  Since this
object  is potentially  a  contaminant,  it is  important  to make  an
estimate  of  the  expected   contribution  from  M31  in  the  larger
$5\arcmin$ sample.  We find that the  halo in our  kinematic survey of
M31,  as seen  through windows  not  affected by  the M31  disk or  by
Galactic stars  \citep{chapman05}, can be approximated  to first order
with  a Gaussian distribution  of dispersion  $99\kms$, centered  at a
mean velocity of  $-300\kms$.  We use the RGB-selected  stars at radii
larger than $5\arcmin$  and with velocities $v <  -300\kms$ (there are
two such  stars) to  normalise this simple  halo model in  the present
field.  A  crude  estimate  of  the expected  contamination  within  a
velocity interval of $\pm 44.6\kms$  of the mean velocity of the inner
sample can  then be made. In  this way we estimate  that $\approx 0.2$
stars   contaminate   the  sample   within   $5\arcmin$,  though   the
uncertainties on this estimate are clearly very large.

There  is therefore  some grounds  to reject  the velocity  outlier at
$1.9\arcmin$. Rejecting  that star in  the maximum-likelihood Gaussian
fit results  in the thick-lined likelihood contours  shown in Figure~4;
the fit  has a mean velocity  and velocity dispersion of  $v = -216\pm
3\kms$  and $\sigma_v =  6.8^{+3.0}_{-2.0}\kms$, respectively.   The lower
panel  of Figure~3  shows the  maximum-likelihood  velocity dispersion
measures, starting at the radius of the inner $1.4\arcmin$ sample, and
moving outwards to $5\arcmin$ adding one datum at a time. For the case
of the thick-lined distribution  we have rejected the velocity outlier
at $1.9\arcmin$.  Evidently, the  first 5 velocity data are consistent
with  being  drawn from  a  population  with  zero intrinsic  velocity
dispersion,  though  beyond  $1.6\arcmin$  the dispersion  appears  to
increase to $\sim 7\kms$. We  note in passing that the present dataset
shows  no  clear radial  velocity  gradient.  The  (maximum-likelihood
fitted)  mean velocity  of the  5 inner  stars within  $1.4\arcmin$ is
$-210.6\pm3\kms$, while  the mean  velocity of the  sample of  5 stars
between  $1.4\arcmin$ and  $5\arcmin$ (with  the outlier  rejected) is
$-220.4\pm4\kms$, so  the velocity difference between  the two samples
is less than $2\sigma$.

\begin{figure}
\psfig{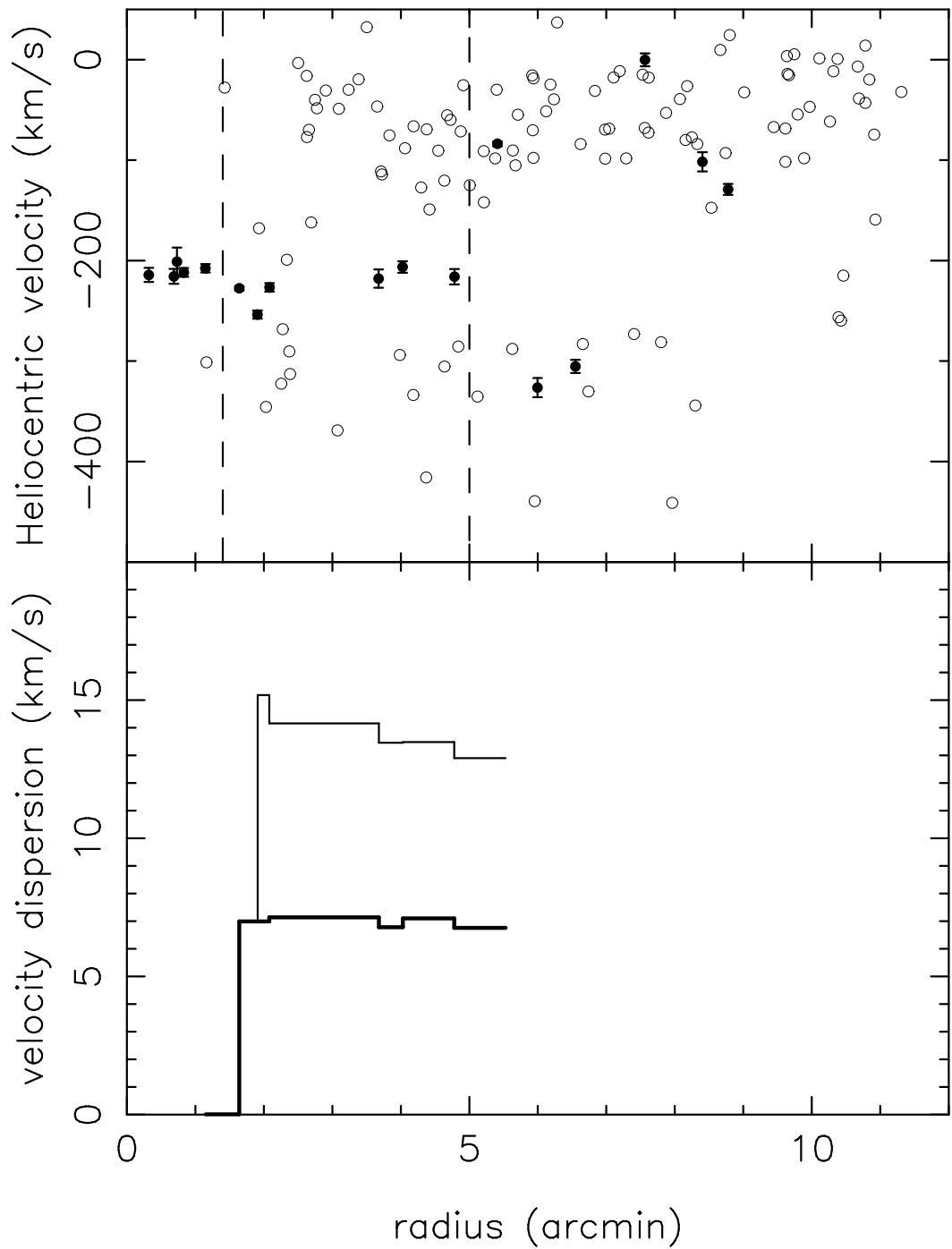}
\caption{Upper panel: the radial velocities of all stars with velocity
uncertainty $<15\kms$  from the 'fiber-hole' mask  centered on \andix.
Solid  circles mark  the data  for  the stars  in the  color-magnitude
selection box of Figure~2; open circles are other targets.  The dashed
lines indicate the radial selection  limits discussed in the text. The
lower panel  shows the derived  maximum-likelihood velocity dispersion
as a  function of radius; the  thin-line corresponds to  all the data,
while  the  thick  line  rejects  one  velocity  outlier  situated  at
$1.9\arcmin$.}
\end{figure}

\begin{figure}
\psfig{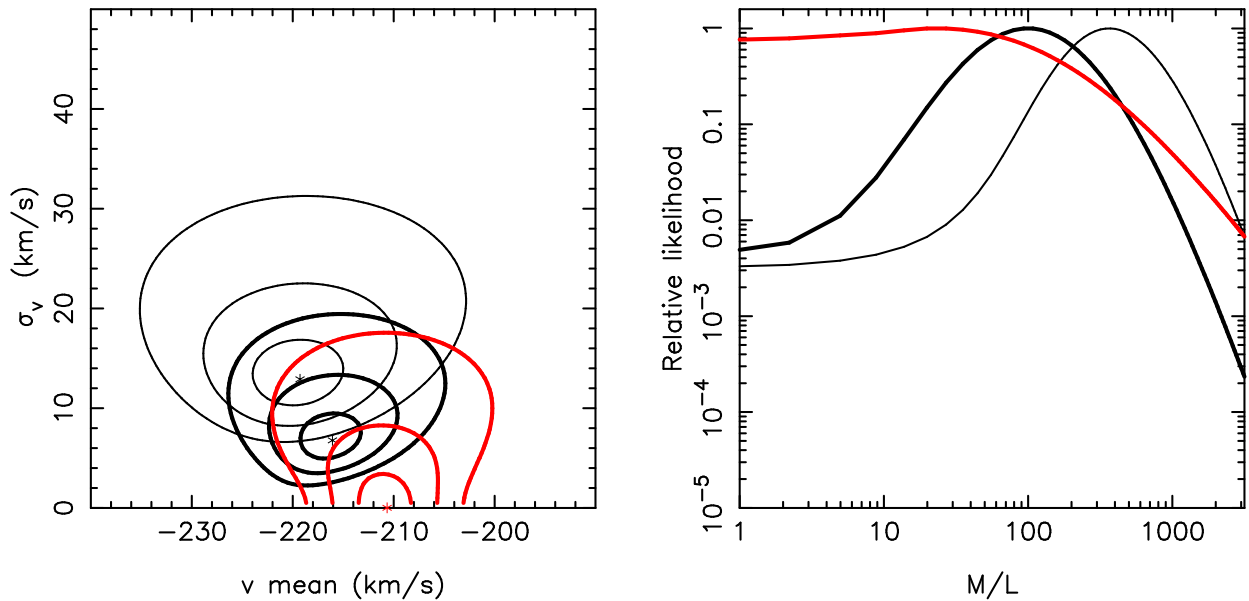}
\caption{The  left-hand  panel   shows  the  likelihood  contours  (at
$1\sigma, 2\sigma,  3\sigma$ intervals) of the mean  and dispersion of
the  \andix\   population,  based  on  radial   velocity  data  within
$5.0\arcmin$ (black  curves) and $1.4\arcmin$ (red  curves).  The thick
black  curves  show  the  effect   of  rejecting  a  single  datum  at
$r=1.9\arcmin$.   The corresponding  relative likelihood  of different
mass to light  ratios are displayed on the  right-hand panel. The thin
black  line is  determined  from the  sample  of 11  stars  with $r  <
5\arcmin$,  while for  the thick  black line  a single  star  has been
rejected.  The red  line shows  the result  for the  $r  < 1.4\arcmin$
sample.}
\end{figure}

The spectra also  allow a measurement of the  metallicity of the dwarf
galaxy using the  CaT equivalent width (EW) technique.   While the S/N
of individual stars are typically too low to yield useful estimates of
the  CaT EWs,  we  proceeded  to measure  the  average metallicity  by
stacking the RGB star spectra of the 5 highest S/N stars of our sample
of 10 stars within $r<5$\arcmin.  We follow as closely as possible the
method  of  \citet{rutledge}, fitting  Moffat  functions  to the  CaII
lines. The average spectrum yields a measurement of the CaT equivalent
width  used to  estimate the  metallicity as  $[Fe/H] =  -2.66  + 0.42
[\Sigma  Ca  -  0.64 (V_{HB}  -  V_{ave})]$,  with  $\Sigma Ca  =  0.5
EW_{\lambda 8498}  + 1.0 EW_{\lambda  8542} + 0.6  EW_{\lambda 8662}$,
$V_{HB}$  being  a surface  gravity  correction  relative  to the  $V$
magnitude  of  the  horizontal  branch, and  $V_{ave}=22.4\pm0.3$  the
average  magnitude of  the \andix\  stars.   We correct  the value  of
$V_{HB}=25.17$ for  M31, measured by  \citet{holland96}, by -0.06~mags
to account  for the difference in  line of sight  distance between M31
and \andix.  We find  $[Fe/H]=-1.5$ (on the \citealt{carretta} scale),
with a large uncertainty of  $\sim 0.3$~mags which is due primarily to
sky-subtraction residuals making it  difficult to define the continuum
level  of  the  spectrum.   This  metallicity  estimate  confirms  the
previous   photometric   estimates    of   $[Fe/H]   \sim   -1.5$   by
\citet{mcconnachie05} and of $[Fe/H] \sim -2$ by \citet{harbeck}.
A metallicity of $[Fe/H] \sim -2$ would be closer to what we would
expect for such a faint dSph, however if the mass were as large as 
suggested by the upper range of our kinematic analysis, 
a metallicity of $[Fe/H] \sim -1.5$ would be reasonable.

\section{A Cold, Dark Matter dominated Accreted Companion to M31?}

The mean radial velocity that we measure ($v=-216 \pm 3 \kms$) implies
that  \andix\ is  almost  certainly  a bound  satellite  to M31.   The
relative radial velocity  from M31 is then only ($\Delta v  = 84 \pm 3
\kms$),  while its separation  from the  M31 center  is only  $\sim 50
\kpc$.   In the  model  of \cite{ibata04},  the  escape velocity  from
$50\kpc$  is   $550\kms$;  if  the  velocity  vector   of  \andix\  is
uncorrelated with the direction vector to the observer, the chances of
measuring a  line of  sight component of  only $84\kms$ if  \andix\ is
unbound would be $< 1$\% (although uncertainty in the line of sight
distance  could  slightly   increase  the  probability).   \andix\  is
therefore most likely bound to M31.

The measured velocity dispersion also  allows us to constrain the dark
matter content of  the small galaxy, if we  assume virial equilibrium.
The  mass to  light ratio  of a  simple spherically  symmetric stellar
system  of  central  surface  brightness $\Sigma_0$,  half  brightness
radius  $r_{hb}$, and  central velocity  dispersion $\sigma_0$  can be
estimated as:
\begin{equation}
M/L =  \eta {{9}\over{2 \pi G}}  {{\sigma^2}\over{ \Sigma_0 r_{hb}}}
\end{equation}
\cite{richstone}, where $\eta$ is  a dimensionless parameter which has
a value close to unity for many structural models. 

For  our sample  of  stars within  $5\arcmin$,  the measured  velocity
dispersion is  not necessarily representative of the  central value of
$\sigma_0$.  
However,   for  most
models of bound stellar systems like those assumed in the core-fitting method,
the  velocity  dispersion decreases with  distance, so
that  our measured dispersion will  be an underestimate of $\sigma_0$
if \andix\ conforms to these models.  
Assuming  that $\sigma_v = \sigma_0$, we
show the relative likelihood of $M/L$ values for our sample selections
on  the right-hand  panel of  Figure~4, where  we have  folded  in the
uncertainties in  the half-light  radius, the velocity  dispersion and
the central  surface brightness.  For  the $5\arcmin$ sample  with the
rejected velocity outlier, we  find that the measured kinematics imply
that \andix\  has a mass to  light ratio 
$M/L  = 93^{+120}_{-50}\msun/\lsun$ (thick
black  line).  In  particular,  with  this sample  we  can reject  the
possibility  that   \andix\  has  $M/L  <   17\msun/\lsun$  with  99\%
confidence. However, this measurement  depends very sensitively on our
adopted  sample;  if we  do  not  reject  the single  most  discrepant
velocity datum we find a higher preferred M/L value (thin black line),
while if we  choose the inner 5 stars alone (red  lines), the mass can
be consistent with zero.

The spectroscopic data we have  presented here have allowed us to make
a preliminary  assessment of  the mass and  nature of  this intriguing
dwarf  galaxy.   While  \andix\  appears  to  be  the  lowest  surface
brightness  galaxy  found  to  date \citep{zucker04a},  and  also  the
faintest   galaxy   known,   our   preferred  mass   estimate   ($\sim
1.6 \times10^{7}\msun$) implies a  substantial dark matter component with
a mass-to-light  ratio $\sim 93$.  For comparison,  applying the same
procedure above  to the case of  the Draco dSph, the  most dark matter
dominated of the Milky Way  satellites, which has $\sigma_0 = 8.5\kms$
\citep{kleyna},  $r_{hb} =  120\pc$ and  $\Sigma_0 =  2.2 \msun/\pc^2$
\citep{irwin95} gives $M/L=91$ (though more detailed dynamical modelling
suggests $M/L = 440 \pm 240\msun/\lsun$ - \citealt{kleyna}).  We point
out that  this analysis assumes  a Gaussian distribution,  while there
are probably too few \andix\ data points to support this assumption.

Although we  cannot rule out  the possibility that tidal  effects have
enhanced  the  extent  and  velocity  dispersion  of  this  satellite,
yielding  a misleadingly high  mass to  light ratio  estimate, \andix\
appears  to be  a  highly  dark matter  dominated  dSph. It  therefore
presents an  intriguing possibility for hierarchical  cold dark matter
models  of structure  formation  -- that  many  dark matter  dominated
dSphs, fainter  even than \andix, and  lying just below  the limits of
detectability, may be present in the halos of giant galaxies like M31,
bringing  models  (e.g.,  \citealt{moore99,  benson02b})  into  better
accord with data.  Nevertheless, given the small size of our sample of
stars  further  accurate kinematic  data  are  needed  to confirm  the
velocity  dispersion  measurement,  while  deep photometry  will  help
ascertain  the extent  to which  tidal  forces may  be perturbing  the
system.

\acknowledgements 
We would like to thank the referee, Steve Majewski, for his thorough comments
on the manuscript which helped improve the paper.
SCC acknowledges support  from  NASA. 
GFL acknowledges support through ARC DP 0343508.
The research
of AMNF has been supported by a Marie Curie Fellowship of the European
Community under contract number HPMF-CT-2002-01758.

\end{document}